\documentclass[a4paper,11pt]{article}
\usepackage{snapshot}
\pdfoutput=1

\usepackage{url}

\usepackage{fullpage}

\usepackage{amsmath,amsthm,amssymb}
\usepackage{xcolor}
\usepackage{graphicx}

\usepackage{tikz}
\usetikzlibrary{decorations.pathreplacing}

\usepackage[position=top]{subfig}
\usepackage{authblk}
\renewcommand\Affilfont{\itshape\small}

\usepackage{trivfloat}
\trivfloat{listing}

\title{magnum.fe: A micromagnetic finite-element simulation code based on FEniCS}

\author[1,2]{Claas Abert\thanks{cabert@physnet.uni-hamburg.de}}
\author[3]{Lukas Exl}
\author[4]{Florian Bruckner}
\author[2]{Andr\'e Drews}
\author[4]{Dieter Suess}
\affil[1]{Fachbereich Mathematik, Universit\"at Hamburg, Bundesstr. 55, D-20146 Hamburg, Germany}
\affil[2]{Institut f\"ur Angewandte Physik und Zentrum f\"ur Mikrostrukturforschung, Universit\"at Hamburg, Jungiusstr. 11, D-20355 Hamburg, Germany}
\affil[3]{University of Applied Sciences, Department of Technology, A-3100 St.Poelten, Austria}
\affil[4]{Institute of Solid State Physics, Vienna University of Technology, A-1040 Vienna, Austria}

\begin{document}
\maketitle

\begin{abstract}
  We have developed a finite-element micromagnetic simulation code based on the FEniCS package called magnum.fe.
  Here we describe the numerical methods that are applied as well as their implementation with FEniCS.
  We apply a transformation method for the solution of the demagnetization-field problem.
  A semi-implicit weak formulation is used for the integration of the Landau-Lifshitz-Gilbert equation.
  Numerical experiments show the validity of simulation results.
  magnum.fe is open source and well documented.
  The broad feature range of the FEniCS package makes magnum.fe a good choice for the implementation of novel micromagnetic finite-element algorithms.

\end{abstract}
  
\newpage
\section{Introduction}
Micromagnetic simulations are an important tool for the computational investigation of ferromagnetic materials.
In recent years they were successfully used to describe magnetic effects ranging from permanent magnets to soft magnetic logic devices to magnetic recording stuctures \cite{zhu_1988,schabes_1991,hertel_2001,klaeui_2005}.
Many methods have been proposed to solve the micromagnetic equations numerically.
Two popular approaches are the finite-difference method combined with the fast Fourier transform (FFT) \cite{berkov_1993,abert_2012,abert_2013} and the finite-element method combined with the boundary-element method \cite{fidler_2000,suess_2002,bruckner_2012}.
Furthermore fast multipole methods \cite{fmm_1997,blue_1991}, nonuniform FFT \cite{alouges_2012} and low-rank tensor methods \cite{exl_2012,exl_fft_2012} have been successfully applied to micromagnetic problems.

Different open-source codes for the finite-difference method \cite{oommf, arne_2011, magnum_homepage} as well as for the finite-element method \cite{nmag,magpar} are available.
Moreover there are a couple of reports on closed source simulation tools \cite{suess_2002,kakay_2010,chang_2011}.
We present the open-source finite-element code magnum.fe that heavily relies on the recently published finite-element software FEniCS \cite{fenics_book} and that solves the dynamic micromagnetic equations with a combination of two weak formulations.

Technically challenging tasks appearing in finite-element computations such as numbering of degrees of freedom, local to global mapping of cell integrals and numerical integration over tetrahedra are handled by FEniCS which offers a variety of finite-element bases including arbitrary order Lagrange elements and produces high performance code for the assembly of system matrices.

The high level of abstraction of FEniCS leads to a very concise code that naturally reflects the underlying numerical algorithms.
This makes magnum.fe an ideal platform for the implementation of new micromagnetic finite-element algorithms.
Implementing alternative weak formulations for certain subproblems can often be done with a few lines of Python.

This paper is structured as follows.
In Sec.~\ref{sec:mm} we briefly present the theory of dynamical micromagnetism.
In Sec.~\ref{sec:numerics} we describe the numerical methods that are implemented in magnum.fe, namely a transformation method for the computation of the demagnetization field and a weak formulation for the integration of the Landau-Lifshitz-Gilbert equation as proposed in \cite{alouges_2008}.
Section~\ref{sec:implementation} gives an overview over the implementation of the algorithms and in Sec.~\ref{sec:experiments} we show the validity of our code by means of numerical experiments.

\section{Micromagnetism}\label{sec:mm}
Magnetization dynamics in the framework of micromagnetism are described by the Landau-Lifshitz-Gilbert equation (LLGE)
\begin{align}
  \partial_t \boldsymbol{m} &= - \gamma ( \boldsymbol{m} \times \boldsymbol{H}_\text{eff} )
                               + \alpha ( \boldsymbol{m} \times \partial_t \boldsymbol{m} )
  \label{eq:llg}
\end{align}
where $\gamma$ is the gyromagnetic ratio and $\alpha \ge 0$ is a phenomenological damping constant that depends on the material.
The magnetization field $\boldsymbol{m}$ is defined on a domain $\Omega$ and assumed to be normalized everywhere
\begin{align}
  |\boldsymbol{m}(\boldsymbol{x})| = 1
  \;,\; \boldsymbol{x} \in \Omega.
  \label{eq:constraint}
\end{align}
This property is obviously preserved by the LLGE \eqref{eq:llg}.
The effective field $\boldsymbol{H}_\text{eff}$ is given by the negative variational derivative of the Gibbs free energy $U(\boldsymbol{m})$ with respect to the magnetization $\boldsymbol{m}$
\begin{equation}
  \boldsymbol{H}_\text{eff}
  = - \frac{1}{\mu_0 M_\text{s}} \frac{\delta U(\boldsymbol{m})}{\delta \boldsymbol{m}}.
\end{equation}
The total effective field $\boldsymbol{H}_\text{eff}$ is the sum of multiple contributions to the Gibbs free energy
\begin{align}
  \boldsymbol{H}_\text{eff} = 
  \boldsymbol{H}_\text{ex} + 
  \boldsymbol{H}_\text{demag} +
  \boldsymbol{H}_\text{zeeman}
  \label{eq:heff}
\end{align}
where $\boldsymbol{H}_\text{ex}$ is the exchange field, $\boldsymbol{H}_\text{demag}$ is the demagnetization field and $\boldsymbol{H}_\text{zeeman}$ is a constant external Zeeman field.

The exchange field $\boldsymbol{H}_\text{ex}$ models the quantummechanical effect of the exchange interaction and is given by
\begin{equation}
  \boldsymbol{H}_\text{ex} = \frac{2 A_\text{ex}}{\mu_0 M_\text{s}} \Delta \boldsymbol{m}
  \label{eqn:exchange}
\end{equation}
where $A_\text{ex}$ is the exchange constant and $M_\text{s}$ is the saturation magnetization.
Including the exchange field in the LLGE \eqref{eq:llg} gives rise to a boundary condition posed on the magnetization $\boldsymbol{m}$
\begin{align}
  \partial_{\boldsymbol{\nu}} \boldsymbol{m}(\boldsymbol{x}) &= 0
  \;,\;\boldsymbol{x} \in \partial{\Omega}
  \label{eq:neumann_condition}
\end{align}
where $\partial_{\boldsymbol{\nu}}$ is the normal derivative.
This boundary condition is often referred to as Brown condition \cite{brown}.

The demagnetization field $\boldsymbol{H}_\text{demag}$ accounts for the magnetic dipole--dipole interaction.
In the absence of electric current, the demagnetization field is curl-free and hence can be expressed as the gradient of a scalar potential $u$
\begin{align}
  \boldsymbol{H}_\text{demag} = - \nabla u.
\end{align}
The magnetic scalar potential $u$ itself is the solution of a Poisson problem
\begin{align}
  \Delta u = M_\text{s} \; \nabla \cdot \boldsymbol{m}.
  \label{eq:poisson}
\end{align}
Boundary conditions for this Poisson problem are given as zero at infinity which is often referred to as open boundary conditions.
With the Green's function of the Laplace operator the solution to \eqref{eq:poisson} can be written in the integral form
\begin{align}
  u(\boldsymbol{x})
  = \frac{M_\text{s}}{4\pi} \int \boldsymbol{m}(\boldsymbol{x'}) \cdot
  \frac{\boldsymbol{x} - \boldsymbol{x'}}{|\boldsymbol{x} - \boldsymbol{x'}|^3} \text{d} \boldsymbol{x'}
  \label{eq:poisson_integral}
\end{align}
which directly fullfills the open boundary conditions, see \cite{abert_2012}.

\section{Numerics}\label{sec:numerics}
\subsection{Demagnetization Field}\label{sec:demag}
\begin{figure}
\centering
\subfloat[]{\includegraphics[width=0.3\textwidth]{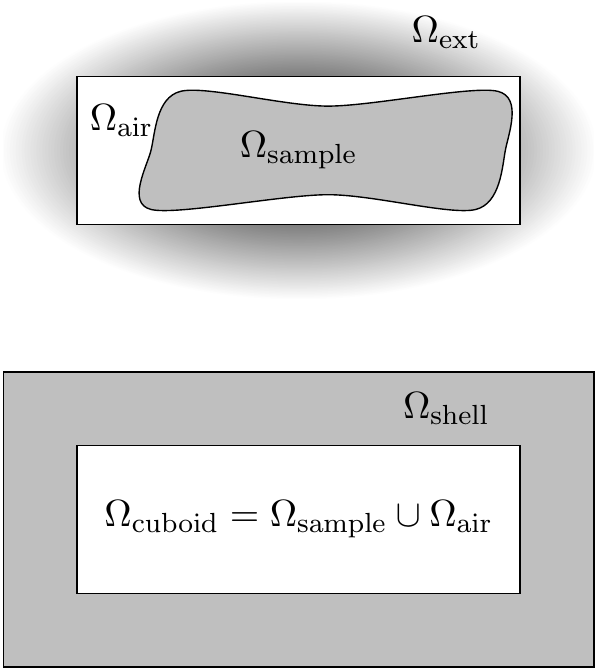} \label{fig:trans_regions}}
\hspace{1cm}
\subfloat[]{\includegraphics[width=0.5\textwidth]{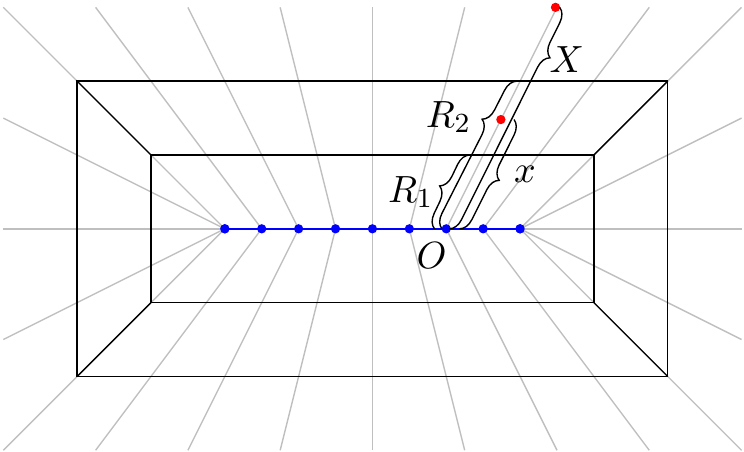} \label{fig:trans_sketch}}
\caption{
The transformation applied for the demagnetization-field computation.
\subref{fig:trans_regions} The regions in transformed and untransformed space.
\subref{fig:trans_sketch} Sketch of the transformation showing the construction of the transformation direction.
The transformation rays and origins marked by gray lines and blue points illustrate the construction of the transformation direction for different points in the shell.
}
\label{fig:trans}
\end{figure}

On a finite region $\Omega$ the Poisson equation \eqref{eq:poisson} with Dirichlet boundary conditions is solved by the weak formulation
\begin{align}
  \int_{\Omega} \nabla u \cdot \nabla v \;\text{d} \boldsymbol{x} =
  M_\text{s} \int_{\Omega_\text{sample}} \boldsymbol{m} \cdot \nabla v \;\text{d} \boldsymbol{x}
  \quad \forall \quad v \in V \subset H^1.
  \label{eq:poisson_weak}
\end{align}
where the boundary conditions are embedded in the function space $V$.
However the problem under consideration has open boundary conditions.
Consequently we would have to carry out the integration in \eqref{eq:poisson_weak} over the whole space which is not possible with the finite-element method.

In order to avoid this restriction we use a method called parallelepipedic shell transformation \cite{brunotte_1992}.
A finite cuboid shell $\Omega_\text{shell}$ is mapped onto the infinite exterior region $\Omega_\text{ext} \in \mathbb{R}^3 \backslash \Omega_\text{cuboid}$ via a bijective transformation $X(x)$
\begin{equation}
  X: \Omega_\text{shell} \rightarrow \Omega_\text{ext}.
  \label{eq:map}
\end{equation}
The integration over $\Omega_\text{ext} = X(\Omega_\text{shell})$ and $\Omega_\text{shell}$ are connected via substitution by
\begin{equation}
  \int_{\Omega_\text{ext}} f(x(X))\, \mathrm{d}X = \int_{\Omega_\text{shell}} f(x) \left|\det(D X(x))\right| \mathrm{d}x.
\end{equation}
When applying this to the left-hand side of \eqref{eq:poisson_weak} the gradients would still be calculated with respect to the shell coordinates $x$ instead of the exterior coordinates $X$.
The gradient with respect to exterior coordinates $X$ is given by
\begin{equation}
  [\nabla_X v](x) = [J^{-1} \nabla_x v](x)
\end{equation}
with $J = D X(x)$ being the Jacobian of the transformation.
Thus the weak formulation with shell transformation reads
\begin{align}
  \int_{\Omega_\text{cuboid}} \nabla u  \cdot \nabla v \;\text{d} \boldsymbol{x} +
  \int_{\Omega_\text{shell}} (\nabla u)^T \boldsymbol{g} \; \nabla v \;\text{d} \boldsymbol{x} =
  M_\text{s} \int_{\Omega_\text{sample}} \boldsymbol{m} \cdot \nabla v \;\text{d} \boldsymbol{x}
  \quad \forall \quad v \in V
  \label{eq:poisson_trans}
\end{align}
with the metric tensor $\boldsymbol g$ given by
\begin{align}
  \boldsymbol{g} = J^{-1,T} J^{-1} \left|\det J\right|.
\end{align}
The Dirichlet boundary condition $u = 0$ on $\partial (\Omega_\text{cuboid} \cup \Omega_\text{shell})$ is embedded in the function space $V$.
The metric tensor $\boldsymbol{g}$ is symmetric positive definite, hence the symmetric bilinear form on the left-hand side of \eqref{eq:poisson_trans} is also positive definite.
Thus, by the right choice of the subspace $V \subset H^1$, problem \eqref{eq:poisson_trans} has a unique solution.
\subsubsection{Choice of Transformation}
We choose the shape of the shell to be cuboid as described in \cite{brunotte_1992}.
The transformation from $\Omega_\text{shell}$ to $\Omega_\text{ext}$ is carried out along rays as sketched in Fig.~\ref{fig:trans_sketch}.
In the simple case of a cubic sample the transformation has a fixed origin and is radial.
However in the general case the origin has to be variable in order to obtain a continuous transformation across shell-patch borders.
The transformation origin in this case moves on the middle plane that is perpendicular to the shortest edge of the cuboid.

In the following the properties of the one-dimensional transformation in the radial directions are discussed.
Obviously there are many possible choices for such one-dimensional transformations that are bijective and fullfill \eqref{eq:map}.
A suitable transformation distorts the basis functions used for discretization in a way that the decay of the potential $u$ may be approximated accurately.
As can be seen in \eqref{eq:poisson_integral} the decay of the potential is $u \propto 1/|\boldsymbol{x}|^2$ in the far-field approximation.
In the following piecewise linear basis functions
\begin{equation}
  \phi(x) = a + b x
\end{equation}
are considered in the untransformed space.
In order to obtain a test function decaying with $1/X^2$ the transformation $X: \Omega_\text{shell} \rightarrow \Omega_\text{ext}$ has to fullfill
\begin{equation}
  x = a' + b' \frac{1}{X^2}
\end{equation}
and thus
\begin{equation}
  X = \sqrt{\frac{b'}{x - a'}}
\end{equation}
Furthermore the transformation $X$ has to map the finite interval $[R_1,R_2]$ to the infinite interval $[R_1,\infty]$, see Fig.~\ref{fig:trans_sketch}.
\begin{align}
  X(R_1) &= \sqrt{ \frac{b'}{R_1 - a'} } = R_1 \\
  1/X(R_2) &= \sqrt{ \frac{R_2 - a'}{b'} } = 0
\end{align}
This immediatly yields
\begin{equation}
  X = R_1 \sqrt{ \frac{R_2-R_1}{R_2-x} }
  \label{eq:transformation_1d}
\end{equation}
as suitable transformation for linear basis functions.

When using higher order polynomials as basis functions we use the transformation
\begin{align}
  X = R_1 \frac{R_2-R_1}{R_2-x}.
  \label{eq:transformation2_1d}
\end{align}
instead of \eqref{eq:transformation_1d}.
This transforms second and third order polynomials like
\begin{align}
  a + bx + cx^2 &\rightarrow a' + b'\frac{1}{X} + c'\frac{1}{X^2} \label{eq:distort_p2}\\
  a + bx + cx^2 + dx^3 &\rightarrow a' + b'\frac{1}{X} + c'\frac{1}{X^2}+ d'\frac{1}{X^3}\label{eq:distort_p3}
\end{align}
which enables a much better approximation of the decaying magnetic potential.

\subsection{Landau-Lifshitz-Gilbert Equation} \label{sec:llg}
We solve the Landau-Lifshitz-Gilbert equation \eqref{eq:llg} with a weak formulation originally proposed by Alouges in \cite{alouges_2008}.
In a first step we set $\boldsymbol{v} = \partial_t \boldsymbol{m}$ and multiply with vector test functions which yields
\begin{align}
  \int_\Omega (\boldsymbol{v} - \alpha \boldsymbol{m} \times \boldsymbol{v}) \cdot \boldsymbol{\phi} \; \text{d} \boldsymbol{x}
  + \int_\Omega \gamma ( \boldsymbol{m} \times \boldsymbol{H}_\text{eff} )   \cdot \boldsymbol{\phi} \; \text{d} \boldsymbol{x}
  = 0
  \quad \forall \quad
  \boldsymbol{\phi} \in V^3.
  \label{eq:llg_weak}
\end{align}
The terms of the right-hand side of the LLGE \eqref{eq:llg} and thus also the left-hand side are perpendicular to the magnetization $\boldsymbol{m}$.
Therefore it is sufficient to test the equation with test functions $\boldsymbol \phi \in T_{\boldsymbol{m}}$ and restrict the solution space of $\boldsymbol{v}$ to $T_{\boldsymbol{m}}$ where $T_{\boldsymbol{m}}$ is the tangent space to the magnetization $\boldsymbol{m}$.
Following Alouges we set $\boldsymbol{\phi} = \boldsymbol{m} \times \boldsymbol{w}$ in \eqref{eq:llg_weak} and restrict the new test functions $\boldsymbol{w}$ to the tangent space $T_{\boldsymbol{m}}$, which yields
\begin{align}
  \int_\Omega ( \alpha \boldsymbol{v} + \boldsymbol{m} \times \boldsymbol{v}) \cdot \boldsymbol{w} \;\text{d}\boldsymbol{x}
  - \gamma \int_\Omega \boldsymbol{H}_\text{eff}(\boldsymbol{m}) \cdot \boldsymbol{w} \;\text{d}\boldsymbol{x}
  &= 0
  \quad \forall \quad
  \boldsymbol{w} \in T_{\boldsymbol{m}}.
  \label{eq:llg_weak_alouges}
\end{align}
This scheme can be extended to an implicit $\theta$-scheme by replacing $\boldsymbol{m}$ with $\boldsymbol{m} + \theta k \boldsymbol{v}$ with the timestep $k$ and $\theta \in [0, 1]$
\begin{align}
  \int_\Omega ( \alpha \boldsymbol{v} + \boldsymbol{m} \times \boldsymbol{v}) \cdot \boldsymbol{w} \;\text{d}\boldsymbol{x}
  - \gamma \int_\Omega \boldsymbol{H}_\text{eff}(\boldsymbol{m} + \theta k \boldsymbol{v}) \cdot \boldsymbol{w} \;\text{d}\boldsymbol{x}
  = 0
  \quad \forall \quad
  \boldsymbol{w} \in T_{\boldsymbol{m}}.
  \label{eq:llg_implicit}
\end{align}
In contrast to \eqref{eq:llg_weak} effective field terms which are linear in the magnetization can be integrated implicitly without breaking the linearity of the scheme.
E.g. considering only the exchange field \eqref{eqn:exchange} and performing integration by parts yields
\begin{align}
  \int_\Omega ( \alpha \boldsymbol{v} + \boldsymbol{m} \times \boldsymbol{v}) \cdot \boldsymbol{w} \;\text{d}\boldsymbol{x}
  + \frac{2 A_\text{ex} \gamma}{\mu_0 M_\text{s}}
    \int_\Omega \nabla (\boldsymbol{m} + \theta k \boldsymbol{v}) \cdot \nabla \boldsymbol{w} \;\text{d}\boldsymbol{x}
  = 0
  \quad \forall \quad
  \boldsymbol{w} \in T_{\boldsymbol{m}}
  \label{eq:llg_exchange}
\end{align}
where the boundary condition \eqref{eq:neumann_condition} has been taken into account.
When using the whole effective field \eqref{eq:heff} it is sufficient to treat $\boldsymbol{H}_\text{ex}$ implicitly in order to get a stable scheme \cite{alouges_2012}.
This is of special practical importance since the calculation of the demagnetization field $\boldsymbol{H}_\text{demag}$ is very time-consuming since the corresponding discretized operators are dense and thus in general not feasible to be computed.
We therefore calculate the demagnetization potential $u$ as described in Sec.~\ref{sec:demag} and treat it only explicitly.
Including $\boldsymbol{H}_\text{ex}$ implicitly and $\boldsymbol{H}_\text{demag}$ and $\boldsymbol{H}_\text{zeeman}$ explicitly yields the weak formulation
\begin{multline}
  \int_\Omega ( \alpha \boldsymbol{v} + \boldsymbol{m} \times \boldsymbol{v}) \cdot \boldsymbol{w} \;\text{d}\boldsymbol{x}
  + \frac{2 A_\text{ex} \gamma}{\mu_0 M_\text{s}}
    \int_\Omega \theta k \nabla \boldsymbol{v} \cdot \nabla \boldsymbol{w} \;\text{d}\boldsymbol{x} \\
  =
  \gamma \int_\Omega (\boldsymbol{H}_\text{zeeman} - \nabla u) \cdot \boldsymbol{w} \;\text{d}\boldsymbol{x}
  - \frac{2 A_\text{ex} \gamma}{\mu_0 M_\text{s}}
  \int_\Omega \nabla \boldsymbol{m} \cdot \nabla \boldsymbol{w} \;\text{d}\boldsymbol{x}
  \quad \forall \quad
  \boldsymbol{w} \in T_{\boldsymbol{m}}
  \label{eq:weak_final}
\end{multline}
which can be written as
\begin{align}
  a(\boldsymbol{v}, \boldsymbol{w}) = L(\boldsymbol{w})
  \quad \forall \quad \boldsymbol{w} \in T_{\boldsymbol{m}}.
  \label{eq:weak_short}
\end{align}
where $a(\boldsymbol{v}, \boldsymbol{w})$ is a bilinear form and $L(\boldsymbol{w})$ is a linear form.
As shown in \cite{alouges_2008} the bilinear form $a(\boldsymbol{v}, \boldsymbol{w})$ can be written as the sum of a skew-symmetric form and a symmetric positive definite form.
Thus the problem \eqref{eq:weak_final} possesses a unique solution $\boldsymbol{v}$.

In order to relieve the tangent-space constraint on the test functions $\boldsymbol{w}$, \eqref{eq:weak_final} has to be supplemented by a term that accounts for the part of $\boldsymbol{w}$ parallel to the magnetization $\boldsymbol{m}$.
Together with the tangent-space constraint for the solution $\boldsymbol{v}$ the system then reads
\begin{align}
  a(\boldsymbol{v}, \boldsymbol{w})
  + \int_\Omega \lambda \; \boldsymbol{m} \cdot \boldsymbol{w} \; \text{d} \boldsymbol{x}
  &=
  L(\boldsymbol{w})
  &\forall& \quad \boldsymbol{w} \in V^3
  \label{eq:weak_final2}
  \\
  \int_\Omega \boldsymbol{v} \cdot \boldsymbol{m} \; \sigma \; \text{d} \boldsymbol{x} &= 0
  &\forall& \quad \sigma \in V.
  \label{eq:tangent_constraint}
\end{align}
If $(\boldsymbol{v}, \lambda)$ is a solution to this system then $\boldsymbol{v}$ solves \eqref{eq:weak_final}.
We discretize \eqref{eq:weak_final2} and the constraint \eqref{eq:tangent_constraint} choosing the same order of finite elements for the scalar field $\lambda$ as for the solution $\boldsymbol{v}$.
This leads to a saddle-point problem of the form
\begin{align}
  \begin{pmatrix}
    A & B \\ B^T & 0
  \end{pmatrix}
  \begin{pmatrix}
    \boldsymbol{v} \\ \lambda
  \end{pmatrix}
  =
  \begin{pmatrix}
    f \\ 0
  \end{pmatrix}
  \label{eq:lin_system}
\end{align}
where $A \in \mathbb{R}^{3N\times3N}$ corresponds to the bilinear form $a(\boldsymbol{v}, \boldsymbol{w})$, $f \in \mathbb{R}^{3N}$ corresponds to the linear form $L(\boldsymbol{w})$ and $B \in \mathbb{R}^{3N\times N}$ and $B^T$ correspond to the integrals in \eqref{eq:weak_final2} and \eqref{eq:tangent_constraint}.
Since $B$ has full rank and $A$ is regular, the system \eqref{eq:lin_system} has a unique solution.

A single integration step is carried out by first solving \eqref{eq:lin_system} for a given magnetization $\boldsymbol{m}(t)$ and then proceed in time by
\begin{align}
  \boldsymbol{m}_i(t+k) = \frac{\boldsymbol{m}_i(t) + k \boldsymbol{v}_i}{|\boldsymbol{m}_i(t) + k \boldsymbol{v}_i|}
  \label{eq:renorm}
\end{align}
where $\boldsymbol{m}_i$ and $\boldsymbol{v}_i$ are the nodal values at node $i$ of the discretized magnetization $\boldsymbol{m}$ and solution $\boldsymbol{v}$.

\section{Implementation}\label{sec:implementation}
Finite element software basically has to solve three sub problems: Mesh generation, system-matrix assembly and solution of the resulting linear systems of equations.
We use Gmsh \cite{gmsh} for mesh generation and FEniCS \cite{fenics_book} for matrix assembly and the solution of linear systems.

\begin{figure}
\centering
\subfloat[]{\includegraphics[width=0.3\textwidth]{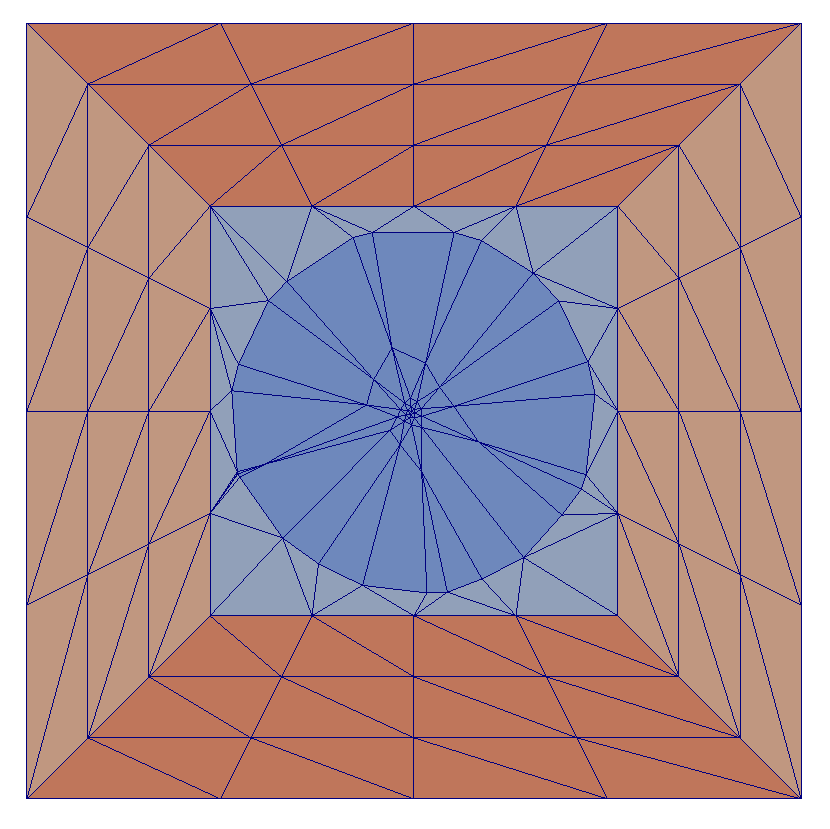} \label{fig:mesh_cut}}
\hspace{2cm}
\subfloat[]{\includegraphics[width=0.3\textwidth]{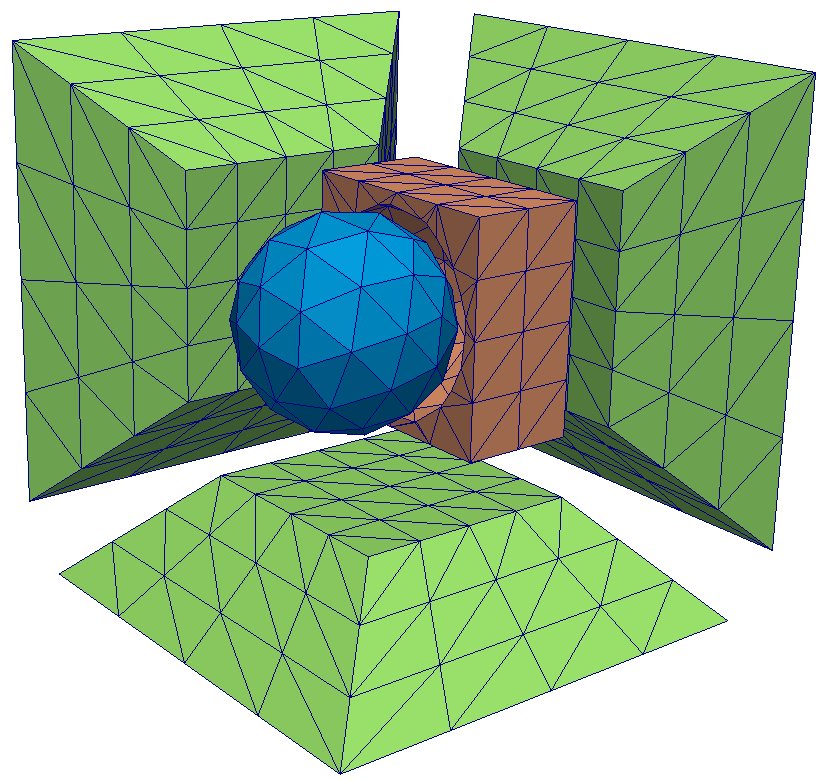} \label{fig:mesh_explosion}}
\caption{
Mesh of a spherical sample including the cuboid shell for the demagnetization-field computation generated with magnum.fe using Gmsh.
\subref{fig:mesh_cut} Cut through the middle plane of the mesh.
\subref{fig:mesh_explosion} Explosion view of the the meshed sphere including the cuboid shell patches.
}
\label{fig:mesh}
\end{figure}
For the automated generation of a suitable mesh for the demagnetization-field computation we use the C++ interface of Gmsh.
magnum.fe is able to produce regular meshes for rectangular samples or alternatively read mesh information from a mesh file, which may implement any format supported by Gmsh.
Then a cuboid shell is wrapped around the sample and meshed with Gmsh, see Fig. \ref{fig:mesh}.
The number of shell layers which largely influences the quality of the demagnetization-field approximation is configurable.

\begin{listing}
  \centering
  \includegraphics[]{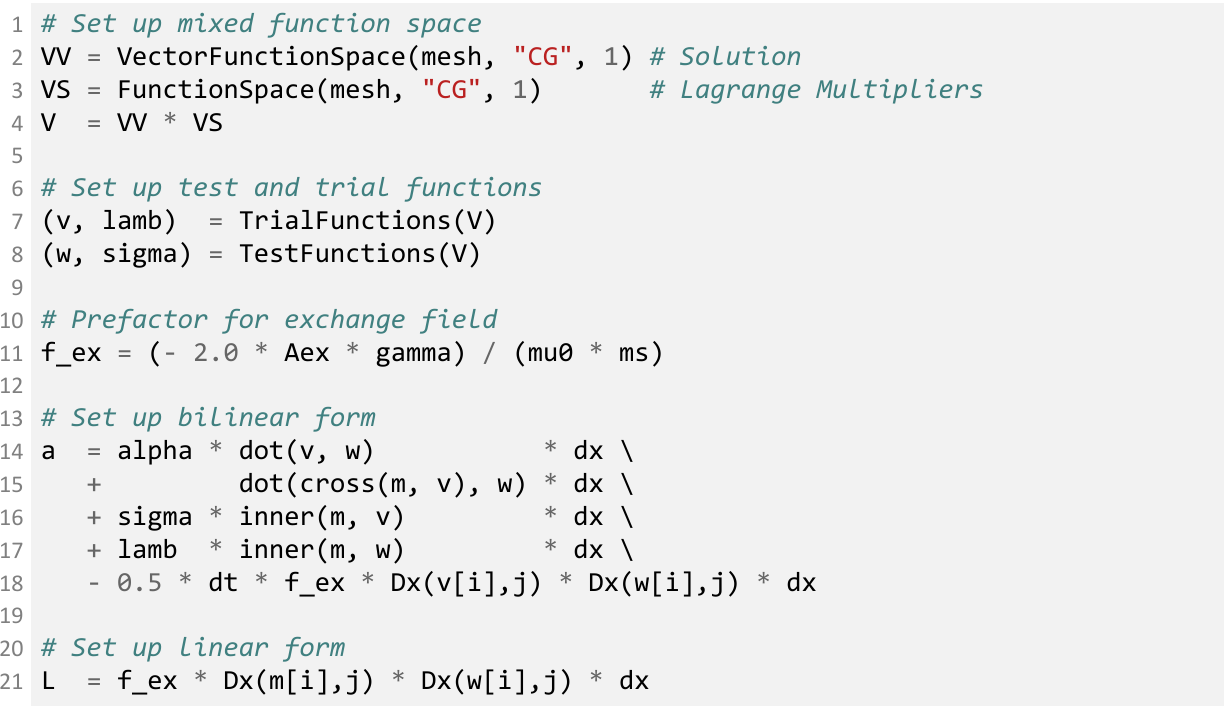}
  \caption{Weak form of the Landau-Lifshitz-Gilbert equation as explained in Sec.~\ref{sec:llg}. A mixed function space is used to include the solution $\boldsymbol v$ and the scalar field $\lambda$ (called lamb here, since lambda is a Python keyword).}
  \label{code:llg}
\end{listing}

\begin{listing}
  \centering
  \includegraphics[]{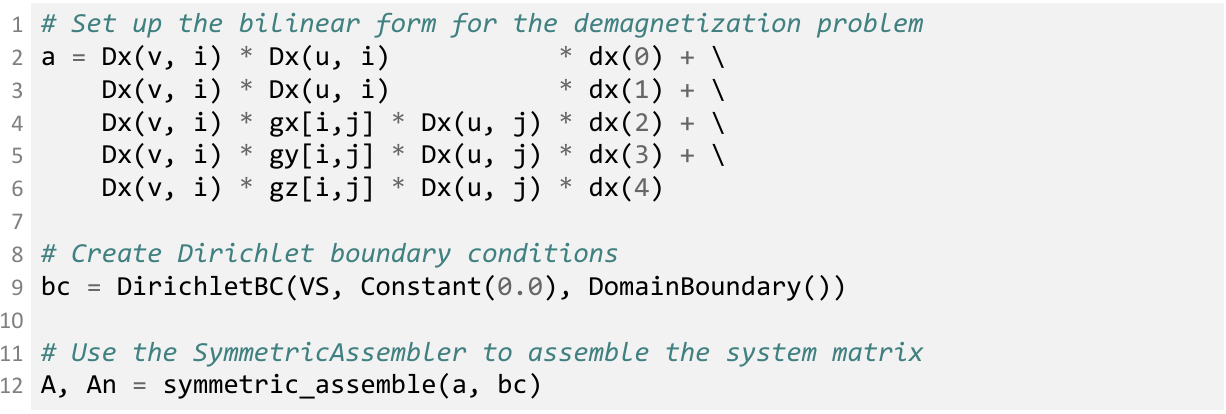}
  \caption{Weak form of the demagnetization-field problem.
  The matrix-valued expressions \texttt{gx}, \texttt{gy} and \texttt{gz} are multiplied using indices \texttt{i}, \texttt{j} and \texttt{k} and integrated over different subdomains of the mesh.}
  \label{code:demag}
\end{listing}

The system-matrix assembly is done with FEniCS which offers a C++ interface as well as a Python interface.
For magnum.fe we mainly use the Python interface which leads to a very concise code that represents the mathematical problem at hand very naturally.
Listing~\ref{code:llg} and \ref{code:demag} show excerpts of the magnum.fe code, namely the definition of the weak formulations \eqref{eq:poisson_trans} and \eqref{eq:weak_final2}-\eqref{eq:tangent_constraint} with the unified form language (UFL)\cite{ufl} defined by FEniCS.
From these form definitions FEniCS creates and compiles a fast C++ code for the matrix assembly via the FEniCS form compiler (FFC)\cite{ffc}.
Thus a high performance is achieved although the actual programming is done in the scripting language Python.

However when hitting the limits of FEniCS it is often not possible to extend the functionality with Python without facing performance issues.
All performance relevant extensions to FEniCS are thus written in C++.
Like FEniCS we use SWIG \cite{swig} to exploit the interface of the extensions in Python.
These extensions include the aforementioned Gmsh interface for the generation of the cuboid shells.
Furthermore an extension for the assembly of matrices and vectors that arise from pointwise calculations of functions was written.
This extension works for $n$th-order Lagrange elements and performs calculations on the (auxiliary-) nodes of the function space.
It is used for the renormalization step \eqref{eq:renorm} which is applied only on the nodes as well as for an alternative implementation of the extended system \eqref{eq:weak_final2}-\eqref{eq:tangent_constraint} where the constraint \eqref{eq:constraint} is restricted to the nodes.

The algorithms presented in Sec.~\ref{sec:numerics} are implemented with Lagrange functions, which are piecewise polynomial and globally continuous.
In case of the demagnetization field the order of the elements is configurable. 
For the solution of the Landau-Lifshitz-Gilbert equation we choose 1st order elements.
Assembly code generated by FEniCS uses Gauss quadrature for integration.
The degree of quadrature is chosen according to the polynomial degree of the integrand and is thus exact for cell-wise polynomial functions.
Analytical expressions are interpolated to a given degree before integration, e.g. we chose the metric tensor $\boldsymbol g$ to be integrated 5th order.

For the solution of the resulting linear systems of equations FEniCS offers interfaces to a variety of open source linear algebra backends.
magnum.fe uses the Trilinos Epetra \cite{trilinos} backend for both the demagnetization problem and the solution of the LLGE.
Since in both cases the resulting system matrices are sparse, iterative Krylov-space methods are applied.
The matrix of the demagnetization problem is symmetric and positive definite as shown in Sec.~\ref{sec:demag}.
Thus a conjugate gradient solver in combination with an algebraic multigrid preconditioner is used.
The saddle-point problem arising from the Landau-Lifshitz-Gilbert equation is solved by an ILU preconditioned GMRES solver.

\section{Numerical Experiments}\label{sec:experiments}
\subsection{Demagnetization Field}
\begin{figure}
  \centering
  \subfloat[]{\includegraphics[width=0.4\textwidth]{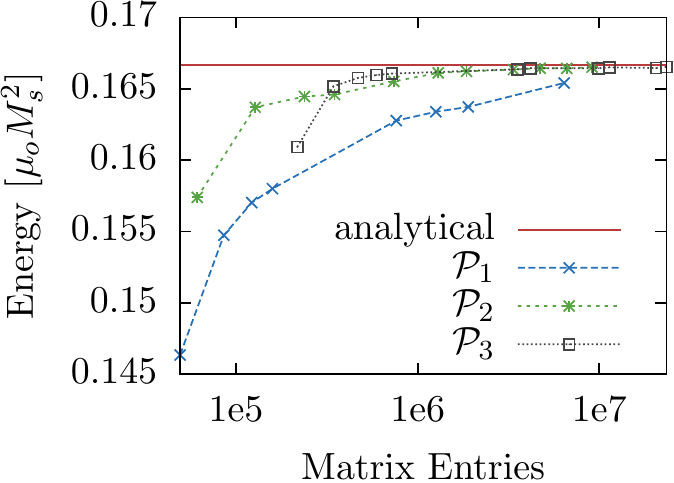} \label{fig:trans_order}}
  \hspace{1cm}
  \subfloat[]{\includegraphics[width=0.4\textwidth]{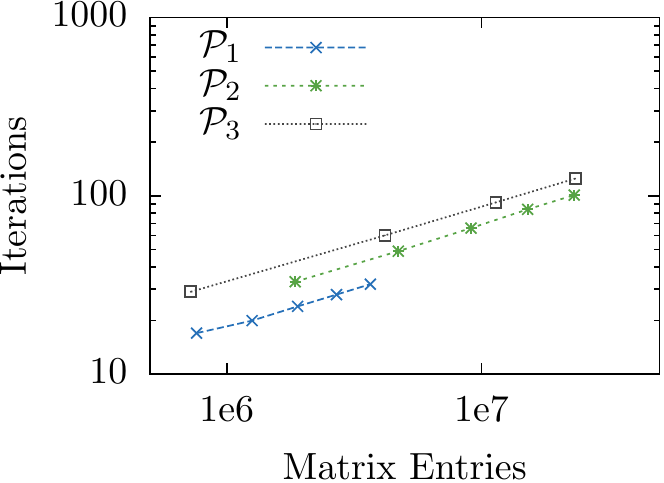}  \label{fig:trans_iter}} \\
  \subfloat[]{\includegraphics[width=0.4\textwidth]{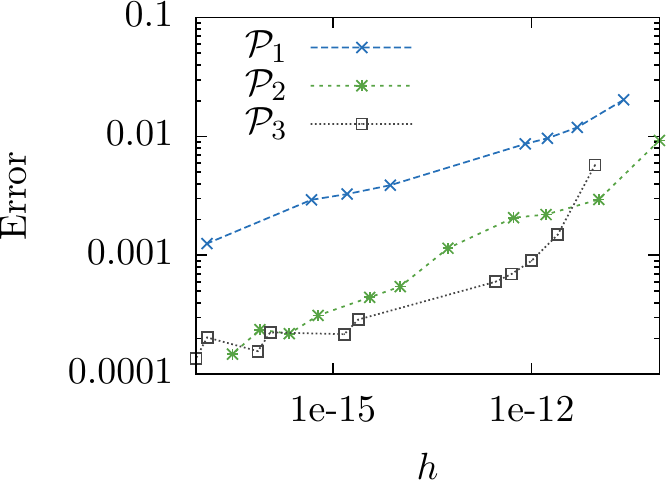}  \label{fig:trans_conv}}
  \hspace{1cm}
  \subfloat[]{
    \hspace{1.2cm}
    \begin{tabular}{c|c}
      degree & rate \\ \hline \hline
      $\mathcal{P}_1$ & $1.58 \pm 0.08$ \\ % 1.57943 \pm 0.07735
      $\mathcal{P}_2$ & $2.34 \pm 0.11$ \\ % 2.33935 \pm 0.1107
      $\mathcal{P}_3$ & $1.81 \pm 0.27$    % 1.80954 \pm 0.2728
    \end{tabular}
    \label{tab:trans_conv}}
  \hspace{1.2cm}

  \caption{
  Demagnetization energy calculations for different order polynomial test and trial functions and different numbers of matrix entries.
  \subref{fig:trans_order} Energy calculations for a homogeneously magnetized unit cube.
  \subref{fig:trans_iter}  Iterations of the Conjugate Gradient solver.
  \subref{fig:trans_conv} and \subref{tab:trans_conv} Convergence rates for different order polynomials.
  }
  \label{fig:trans_plots}
\end{figure}

For validation and comparison of the demagnetization-field algorithm of different order as presented in Sec.~\ref{sec:demag} we choose a homogeneously magnetized unit cube.
The energy of this system can be computed as
\begin{align}
  E = \frac{1}{2} \int_\Omega \nabla u \cdot \boldsymbol{m} \; \text{d} \boldsymbol{x}
\end{align}
which is $1/6$ if $\boldsymbol{m} = (0,0,1)$.
Figure~\ref{fig:trans_plots} shows results for different polynomial degree of the basis functions and different number of non-zero system-matrix entries.
The latter was chosen as measure since a matrix--vector multiplication is linear in this size and so is the iterative solution of the associated linear system.
Also the storage requirements are linear in the number of non-zero matrix entries.

Figure~\ref{fig:trans_order} shows the results of the energy calculations.
The 2nd and 3rd order method perform clearly better than the 1st order method for the same number of matrix entries.
This is a consequence of the additional $1/X$ term that the higher order elements provide in contrast to the 1st order elements, see \eqref{eq:distort_p2} and \eqref{eq:distort_p3}.

Figure~\ref{fig:trans_iter} shows the number of iterations needed for the iterative solution of the linear system of equations.
Together with the numerical complexity of a single matrix--vector multiplication, the number of iterations gives the over-all complexity of the demagnetization-field algorithm.
The log--log plot yields a linear dependence with slope $\approx 0.4$, resulting in an over-all complexity of $\mathcal{O}(N^{1.4})$ for the demagnetization-field algorithm.

Finally Fig.~\ref{fig:trans_conv} and Tab.~\ref{tab:trans_conv} show the convergence rates of the algorithm.
In order to account for auxiliary nodes for the higher order methods the discretization $h$ is not taken from the mesh, but estimated by $N^{-1/3}$ where $N$ is the number of degrees of freedom, i.e. the total number of nodes and auxiliary nodes.
Again the 1st order method shows the poorest performance.
The 2nd and 3rd order methods have convergence rates of approximately 2 which is expected from other demagnetization methods, see \cite{abert_2013}.

For further numerical experiments with the presented demagnetization-field method and comparsion to other recently developed methods see \cite{abert_2013}.

\subsection{Landau-Lifshitz Equation}
\begin{figure}
  \centering
  \includegraphics[width=0.8\textwidth]{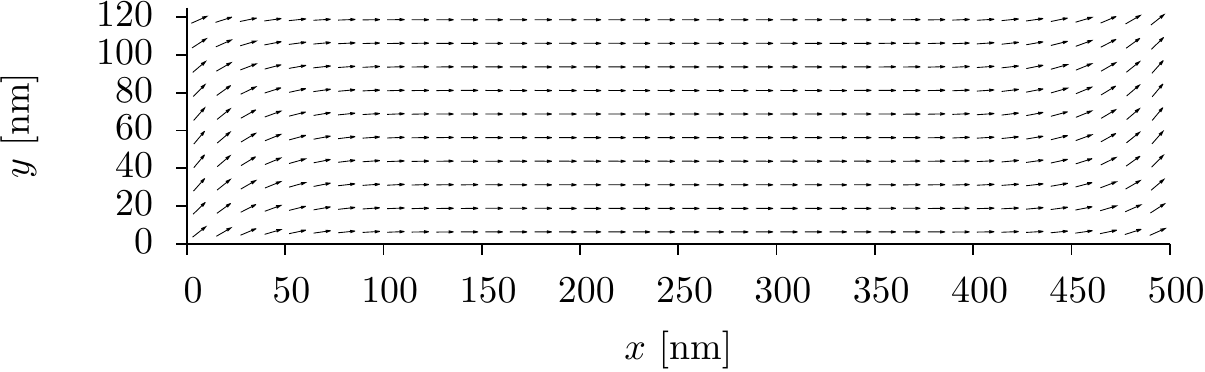}
  \caption{In-plane magnetization configuration of a so-called s-state in a rectangular thin film of the size $500 \times 125 \times 3$nm$^3$ with the material parameters of permalloy.}
  \label{fig:s-state}
\end{figure}

\begin{figure}
  \centering
  \includegraphics[width=0.6\textwidth]{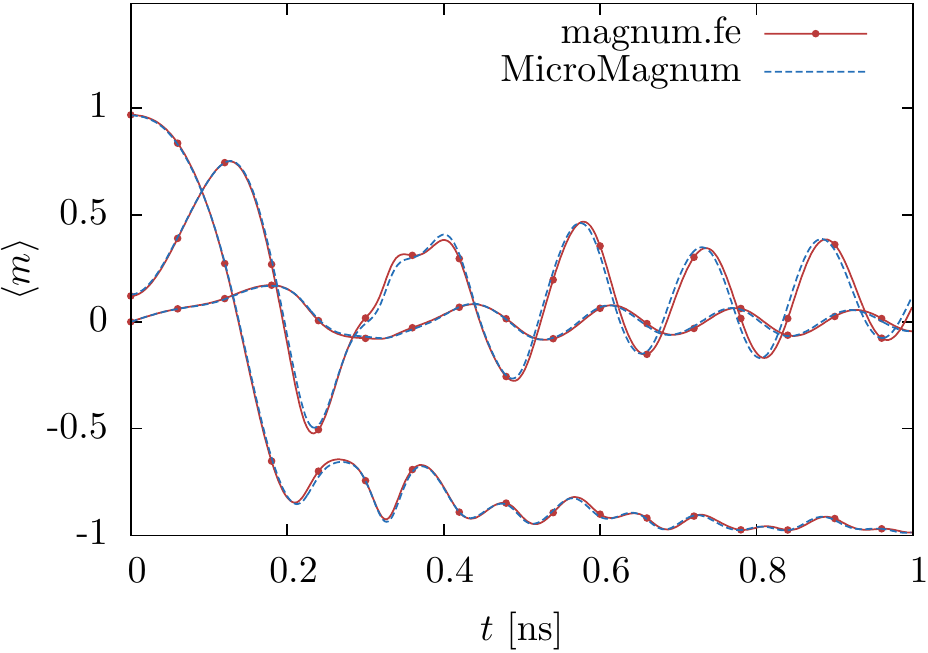}
  \caption{
  Time evolution of the averaged magnetization for the standard problem \#4.
  The results of magnum.fe are compared to that of the finite-difference package MicroMagnum.
  The different branches correspond to the components of the averaged magnetization.
  }
  \label{fig:sp4}
\end{figure}

The method for the integration of the Landau-Lifshitz-Gilbert equation is validated with the standard problem \#4 proposed by the Micromagnetic Modeling Activity Group $\mu$MAG \cite{mumag4}.
A rectangular sample of the size $500 \times 125 \times 3$nm$^3$ is relaxed in a so-called s-state with the bulk magnetization pointing in the $x$-direction, see Fig.~\ref{fig:s-state}.
The material parameters of the sample are chosen similar to those of Permalloy
\begin{align}
  A_\text{ex} &= 1.3 \cdot 10^{-11} \text{J/m} \\
  M_\text{s}  &= 8.0 \cdot 10^{5}   \text{A/m} \\
  \alpha      &= 0.2.
\end{align}
Then, in addition to the exchange field and the demagnetization field, a homogeneous external Zeeman field $\boldsymbol{H}_\text{zeeman} = (-24.6, 4.3, 0)$mT is applied which results in a switching process.
Figure~\ref{fig:sp4} shows the evolution of the averaged magnetization components in time as calculated by magnum.fe compared to the results of the finite-difference code MicroMagnum \cite{magnum_homepage}.


\section{Conclusion and Outlook}
We present the open-source micromagnetic software magnum.fe.
magnum.fe is a complete three-dimensional finite-element code, which computes magnetization dynamics with a combination of two linear weak formulations.
It is written in C++ and Python and uses the finite-element package FEniCS \cite{fenics_book}.
The correctness of the code is demonstrated by a number of numerical experiments including the $\mu$MAG standard problem \#4.

Due to the multitude of features and the high level of abstraction of FEniCS, magnum.fe is well suited for the implementation of novel finite-element algorithms.
magnum.fe itself is well documented and unit tested and is freely available at github \cite{magnumfe_github}.

We plan to implement alternative demagnetization-field algorithms as well as integration schemes for the Landau-Lifshitz-Gilbert equation.
Contributions to magnum.fe are very welcome.

\section*{Acknowledgements}
We thank Michael Hinze and Guido Meier for fruitful discussions.
Financial support by
the Deutsche Forschungsgemeinschaft via the Graduiertenkolleg 1286 ``Functional Metal-Semiconductor Hybrid Systems'',
the Austrian Science Fund (FWF, project SFB-ViCoM F4112-N13)
, and the Sonderforschungsbereich 668 ``Magnetism from the single atom to the nanostructure''
is gratefully acknowledged.

\bibliographystyle{ieeetr}
\bibliography{refs}
\end{document}